\documentclass[twoside,final]{dis04}
\def\runauthor{Andrea Banfi}
\def\shorttitle{Dijet rates with symmetric transverse energy cuts}
\usepackage{epsfig}
\usepackage{amsmath}

\begin{document}
\title{DIJET RATES WITH SYMMETRIC CUTS}

\author{ANDREA BANFI}

\address{NIKHEF Theory group\\
P.O. Box 41882 1009 DB Amsterdam, the Netherlands\\
E-mail: andrea.banfi@nikhef.nl}

\maketitle

\abstracts{ We show that a resummation of infrared logarithms is
  needed to obtain a sensible theoretical description of dijet rates
  when symmetric cuts are applied to the transverse energies of both
  jets.  We also present the next-to-leading logarithmic
  (NLL) resummation we carried out for DIS production of two jets
  selected with the cone algorithm.  }
  
\section{Introduction}
It was observed some time ago~\cite{KK,FrixRid} that next-to-leading
order (NLO) QCD calculations are not able to describe dijet rates
measured experimentally in the whole of the dijet phase
space~\cite{ZEUS,H1}. In particular in~\cite{FrixRid} a study was
performed of the cross section $\sigma(\Delta)$ for producing two jets
with transverse energies $E_{t1}>E_{t2}$, $E_{t2}>E_{\min}$ and
$E_{t1}>E_{\min}+\Delta$.  There it was observed that $\sigma(\Delta)$
at NLO was finite for $\Delta=0$ (symmetric cuts), but in the vicinity
of that point the slope $\sigma'(\Delta)$ became unphysically positive
and infinite.  This pathological behaviour can be attributed to the
fact that incomplete real-virtual cancellations give rise to large
logarithms $\ln Q/\Delta$ (with $Q$ the hard scale of the process) at
all orders in the perturbative expansion for $\sigma'(\Delta)$.
However, an all-order resummation of such logarithms restores the
correct physical behaviour for the slope, as will be discussed
in the following.

\section{Resummation of the jet transverse energy difference}
\label{sec:resummation}
To understand the physical origin of the divergent behaviour of
$\sigma'(\Delta)$ we first observe that this quantity can be
 related to the differential cross section in the
transverse energy of the highest $E_t$ jet:
\begin{equation}
  \label{eq:slope}
  \sigma'(\Delta) = -\frac{d\sigma}{d E_{t1}}(E_{t1}=E_{\min}+\Delta)\,.
\end{equation}
Being the opposite of a physical cross section, the slope has to be
always negative, as is observed experimentally~\cite{ZEUS,H1}.

However, if we emit an arbitrary number of soft gluons $k_i$, and
define $E_{t,\mathrm{jet}} = |\sum_{i\in \mathrm{jet}}\vec k_{ti}|$,
in a recombination scheme that adds transverse momenta vectorially, we
have, from momentum conservation,
\begin{equation}
  \label{eq:etdif}
  E_{t1}-E_{t2} = |\sum_{i \notin \mathrm{jets}} k_{xi}|\,,
\end{equation}
where $k_{x}$ is the component of $\vec k_{t}$ parallel to the jet
axis (which can be taken as the direction of any of the two hard
jets). The definition in eq.~(\ref{eq:slope}) implicitly imposes the
constraint $  |\sum_{i \notin \mathrm{jets}} k_{xi}| <
\Delta$ on emitted particle transverse momenta.
Such a veto on soft radiation 
is known to give rise to terms $\alpha_s^n \ln^m Q/\Delta$, which for
$\Delta \ll Q$ diverge making $\sigma'(\Delta)$ become infinite, as was
first observed in~\cite{FrixRid}.

We then consider the probability $S(q_x)$ of producing an arbitrary
number of soft/collinear (SC) partons such that $q_x = \sum_{i \notin
  \mathrm{jets}} k_{xi}$. For $\Delta \ll Q$, SC emissions factorise
from the leading order value for the slope $\sigma'_0(\Delta)$ as follows:
\begin{equation}
  \label{eq:slope-resum}
  \sigma'(\Delta)\simeq\sigma'_0(\Delta)\cdot\Sigma(\Delta)\,,\qquad
  \Sigma(\Delta) \equiv \int \!dq_x \,S(q_x)\, \Theta(\Delta -|q_x|)\,.
\end{equation}
Assuming that multiple SC emissions are distributed according to a
`random walk'~\cite{PP} around $q_x =0$ leads to $\Sigma(\Delta)\sim
\Delta/Q$ for small $\Delta$. The resulting slope is finite and
negative defined as expected. QCD radiation is a particular form of
random walk, giving rise to the following expression for
$\Sigma(\Delta)$ (see~\cite{symcuts}):
\begin{equation}
  \label{eq:Sigma}
  \Sigma(\Delta) = \frac{2}{\pi}\int_0^{\infty} 
  \frac{db}{b} \sin(b\Delta) e^{-R(b)}\,,
\end{equation}
where the positive defined `radiator' $R(b)$ acts as an effective
cutoff on the $b$ integral. Its {\em perturbative} expansion can be
reorganised as follows:
\begin{equation}
  \label{eq:logarithms}
  -R(b) = Lg_1(\alpha_s L) + g_2(\alpha_s L) + \alpha_s g_3(\alpha_s
  L)+\dots\,,
  \qquad
  L = \ln b\,,
\end{equation}
where $Lg_1(\alpha_s L)$ resums leading logarithms ($\alpha_s^n L^{n+1}$, LL),
$g_2(\alpha_s L)$ resums  next-to-leading logarithms ($\alpha_s^n L^n$, NLL)
and so on. The NLL expression for $R(b)$ reads
\begin{equation}
  \label{eq:R-NLL}
  R(b) = R_\mathrm{inc}(b)+R_\mathrm{soft}(b)+R_\mathrm{jet}(b)+
  R_\mathrm{ng}(b)\,.
\end{equation}
$R_\mathrm{inc}(b)$ collects contributions from radiation that is soft
and collinear to the incoming parton(s). Its expression is analogous
to the well-known Sudakov exponent in Drell-Yan transverse momentum
distribution~\cite{DYkt}. The term $R_\mathrm{soft}(b)$ accounts for
coherence of QCD radiation arising from interference among soft gluons
at large angles.  Both $R_\mathrm{inc}(b)$ and $R_\mathrm{soft}(b)$
are {\em universal}, in the sense that they do not depend on the
details of the jet algorithm.  However, $R_\mathrm{soft}(b)$ should be
corrected by taking into account the fact that not all soft gluons
contribute to $E_{t1}-E_{t2}$, but only those outside the jets. The
correction is provided by the additional terms $R_\mathrm{jet}(b)$ and
$R_\mathrm{ng}(b)$, both of which depend on the jet algorithm.
$R_\mathrm{jet}(b)$, as well as $R_\mathrm{inc}(b)$ and
$R_\mathrm{soft}(b)$, can be computed by simply considering the
emission of a single SC gluon (including its multiple SC splittings)
and exponentiating the result. However, NL logarithms arise also when
soft gluons inside the jets emit coherently a relatively softer gluon
outside. Such contributions are typical of non-global observables
~\cite{DassalNG1} like $E_{t1}-E_{t2}$, and are embodied in
$R_\mathrm{ng}(b)$. Their resummed expression, known only in the
large-$N_c$ limit, has been already computed for two outgoing jets in
$e^+e^-$ annihilation, both in the cone algorithm~\cite{DassalNG2} and
in the inclusive $k_t$ algorithm~\cite{AppSey}.

\begin{figure}[!thb]
\begin{center}
\epsfig{file=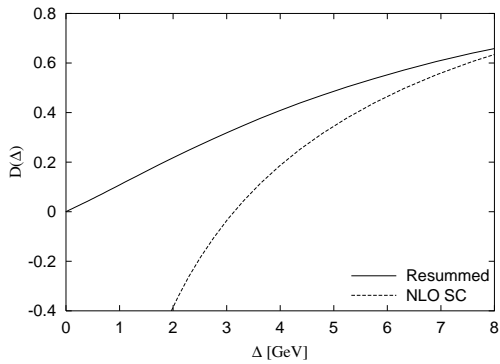, width=.55\textwidth}
\caption[*]{ The ratio
  $D(\Delta)=\sigma'(\Delta)/\sigma'_0(\Delta)$. See \cite{symcuts}
  for details.
\label{fig:ratio}
}
\end{center}
\end{figure}

We now discuss the phenomenological impact of such a resummation for
dijet production in DIS. The explicit expression for $R(b)$ in this
specific case can be found in~\cite{symcuts}.  Here we present
numerical results corresponding to $Q= 20$GeV, $x_B=0.01$, where $-Q^2$
is the virtuality of the photon and $x_B$ the usual Bjorken variable.
Jets are selected with the variant of the cone algorithm proposed
in~\cite{KidSter} with an opening angle $\delta = 0.3$, minimum
transverse energy $E_{\min} = 10$GeV and rapidity $|\eta| < 1$ in
the Breit frame.  In fig.~\ref{fig:ratio} we plot the ratio
$D(\Delta)\equiv\sigma'(\Delta)/\sigma'_0(\Delta)$, with the slope
$\sigma'(\Delta)$ computed at NLO in the SC approximation (lower
curve) and NLL resummed (upper curve). As can be seen, while the NLO
curve diverges for $\Delta \ll Q$, the resummed curve stays finite,
vanishing as $\Delta/Q$.
The numerical analysis of~\cite{symcuts} shows also that the impact of the non-global
piece $R_\mathrm{ng}(b)$ turns out to be negligible in the whole range
of $\Delta$ values considered. This is consoling in view of the fact
that its expression is known only in the large-$N_c$ limit.

\section{Obtaining the total dijet rate}
\label{sec:total-rate}
After a resummation for the slope $\sigma'(\Delta)$ has been computed,
and after matching to fixed order calculations, one could obtain the
total dijet rate $\sigma(0)$ by simply integrating $\sigma'(\Delta)$
from the maximum kinematically allowed value $\Delta_{\max}$ back to
$\Delta = 0$.  However, resumming the distribution in the transverse
energy difference is not the only way to compute $\sigma(0)$. 
One can envisage the possibility of studying other observables, 
which we will discuss shortly.

First of all one can observe that in the massless $E_0$ recombination
scheme $E_{t1}-E_{t2}$ takes contribution also from the invariant
masses of the two outgoing jets. This makes the observable global, and
therefore within the scope of the automated resummation program
\textsc{caesar}~\cite{caesar}. However, one should be aware of the
fact that, since a mechanism that keeps $E_{t1}-E_{t2}$ small is
transverse momentum cancellation (see eq.~(\ref{eq:etdif})), one
should expect a divergence in the resummed distribution provided by
\textsc{caesar} at a critical value $\Delta_c$ (see~\cite{caesar} for
a discussion on this point).

Another possibility would be to consider any observable $V$ that
vanishes in the limit of two outgoing partons ($E_{t1}-E_{t2}$ is only
an example of such an observable). One could then resum the
differential distribution $d\sigma(V)/dV$ and integrate it
from $V_{\max}$ to $V=0$, thus obtaining again the total dijet rate.
This has been observed already in~\cite{FrixRid} for the
case of the azimuthal angle $\phi$ between the jets, where $V=
\pi-\phi$ was the quantity to be resummed.

\section{Outlook}
\label{sec:the-end}
What we have done so far~\cite{symcuts} shows that we have a clear
understanding of the physics underlying dijet rates with symmetric
$E_t$ cuts.  As a mandatory test of our ideas, we plan to extend our
calculation in DIS to the $k_t$ algorithm~\cite{progress}, so that we
can compare our predictions with existing data~\cite{ZEUS,H1}. Since
the data have very small errors, such a comparison would also offer a test
of QCD predictions in a three-jet environment, complementary to the
event-shape measurements proposed for instance in~\cite{3jet}, and, as we
believe, less affected by non-perturbative hadronisation corrections.
Further developments should include the extension of our results to
other hard processes. In particular a study of dijet rates in
photoproduction could be exploited to better constrain the gluon
distribution $g(x)$ at moderately large $x$, which should be important
for the LHC~\cite{LHC}.

\section*{Acknowledgements}
It was a pleasure for me to carry out this work together with Mrinal
Dasgupta.  I am then grateful to him for this and also for having
invited me to participate in the DIS 2004 workshop.

\end{document}